\documentstyle[aps,prl,multicol,epsfig,cite,enumerate,verbatim]{revtex}
\draft \topmargin -1 cm
\newcommand{\expon}{\rm e}
\newcommand{\U}{{\cal U}}
\newcommand{\beq}{\begin{equation}}
\newcommand{\eeq}{\end{equation}}
\newcommand{\bseq}{\begin{eqnarray}}
\newcommand{\eseq}{\end{eqnarray}}
\newcommand{\dis}{\displaystyle}

\begin{document}
\title{Searching a Database under Decoherence}
\author{JingLing Chen$^{a,c}$,  Dagomir Kaszlikowski $^{b,c}$
, L.. C. Kwek$^{c d}$
 and C. H. Oh$^{c}$
}
\address{$^a$ Laboratory of Computational Physics, \\
Institute of Applied Physics and Computational Mathematics, \\P.O.
Box 8009(26), Beijing 100088, People's Republic of China \\ $^b$
Instytut Fizyki Do\'{s}wiadczalnej, \\ Uniwersytet
Gda\'{n}ski, PL-80-952
\\ $^c$ Department of Physics, Faculty of Science, National
University of Singapore, \\ Lower Kent Ridge, Singapore 119260,
Republic of Singapore. \\ $^d$ National Institute of Education,
Nanyang Technological University,
\\1 Nanyang Walk, Singapore 639798}
\maketitle {\small
\begin{abstract}
We report on the effects of a simple decoherence model on the
quantum search algorithm. Despite its simplicity, the decoherence
model is an instructive model that can genuinely imitate realistic
noisy environment in many situations. As one would expect, as the
size of the database gets larger, the effects of decoherence on
the efficiency of the quantum search algorithm cannot be ignored.
Moreover, with decoherence, it may not be useful to iterate beyond
the first maxima in the probability distribution of the search
entry. Surprisingly, we also find that the number of iterations
for maximum probability of the search entry reduces with
decoherence.
\end{abstract}
 }

\begin{multicols}{2}
\section { Introduction}

In many computations, the search algorithm is one of the most
time-consuming activity of the computer program \cite{knuth}. With
the rapid proliferation of the internet and the increasing need to
retrieve information from this ever expanding internet network, it
is crucial to consider alternatives to current search algorithms.
Recently, quantum algorithms have been shown to reduce hitherto
computationally difficult problems for which there are no known
polynomial time algorithm into a tractable problem involving
polynomial time\cite{shor}. This problem involves factoring a
large composite integer into its prime factors. Besides prime
number factorization, it has also been shown that quantum
algorithms like the quantum search algorithm could substantially
improve the searching process for a
database\cite{grover1,grover2,grover3} despite only a square root
speed-up.

Indeed, in recent years, this immense potential in quantum
algorithm has spurred a fecundity of ideas on the actual
implementation of a quantum computer. However, the prospect of
actually building one in the next decade remains taunting and
seemingly insurmountable with current technology due primarily
several reasons, like decoherence and scalability. Unlike
classical bits, quantum bits (qubits) are highly susceptible to
collapse due to the difficulty of isolating the quantum mechanical
systems from its environment. Such decoherence inevitably leads to
a loss of information within the system.  Thus it is necessary to
consider fault tolerant computation through quantum error
correction \cite{preskill} or more recently using decoherence-free
subspaces\cite{lidar,almut}.

Recently, some interesting works have been done to consider the
robustness of the quantum search algorithm under a noisy
environment \cite{pablo,sun}. In ref. \cite{pablo}, they modeled
the decoherence of the search algorithm using a stochastic white
noise. In this paper, by considering a specific decoherence model,
we obtain analytic expression for the probability of the search
item in a quantum search algorithm after a certain number of
iterations. Our analytic expression can certainly facilitate the
study of the behavior and efficiency of the search algorithm
within the model.

This paper is organized as follows.  In section \ref{grover}, we
briefly describe the Grover's search algorithm under a noise-free
environment. In section \ref{search}, we describe our decoherence
model and in section \ref{noisysrch} derive an explicit analytical
formula for the probability of the search item after a certain
number of iterations within the model. In section \ref{discuss},
we consider the robustness of the search algorithm under the
decoherence model. Finally in section \ref{conclude}, we summarize
our results and discuss some implications of the model.

\section{Grover's Search}\label{grover}

In a series of seminal papers, Grover
\cite{grover1,grover2,grover3} considered a quantum algorithm that
could achieve a speed-up in the computational implementation with
only $O(\sqrt{N})$ for a large structured database with $N$
records. This algorithm compares favorably with the classical
result which can only execute a search with $O(N)$ efficiency.
Moreover, it has been shown that Grover's algorithm is optimal
\cite{bennett}.

Grover's search algorithm can be summarized neatly into the
following main steps: (i) Initialization of the system into a
superposition of states; (ii) Subjecting the system to a hashing
function, C(S), represented by a unitary operator, $U$, given by
\begin{equation}
U = \left( \begin{array}{ccccc} -1& 0 & 0 & \cdots & 0 \\ 0 & 1 &
0 & \cdots & 0 \\ \vdots & 0 & 1 & \cdots & \vdots \\ \vdots & 0 &
\vdots & \vdots & \vdots \\ 0& 0 & 0 & \cdots  & 1
\end{array} \right)
\end{equation}
(assuming that the first entry satisfies the search criteria and
therefore undergoes a rotation of $\pi$ radians)
 followed by a
diffusion matrix, $D$, defined by
\begin{equation}
D_{ij} = -\delta_{ij} + \frac{2}{N}
\end{equation}
for $O(\sqrt{N})$ iterations; (iii) Measuring the resulting state.
The heart of the process therefore hinges significantly on the on
step (ii) and the $O(\sqrt{N})$ iterations of the matrix $DU
\equiv S $. In this paper, we look closely into this important
step and scrutinized the ideas behind the efficiency.

Before proceeding further, we first consider the eigenvalues of
the matrix $S$. It is not difficult to show that the eigenvalues
of $S$ all lie on the locus $|z| = 1$, unit circle, on the complex
Argand plane and are explicitly $\displaystyle \{\underbrace{-1,
\cdots, -1}_{(N-2)}, \eta, \eta^\ast \}$ where the root $\eta$ and
$\eta^\ast$ satisfy the equation $\displaystyle z^2 -
\frac{2(N-2)}{N}z + 1 = 0$. Indeed, it can be further shown that
if the matrix $S$ is diagonalized as $P \Lambda P^{-1}$, with
$\Lambda = \mbox{\rm Diag} [ \underbrace{-1, \cdots, -1}_{(N-2)},
\eta, \eta^\ast ]$ and
\begin{equation}\label{rt}
\eta = \frac{1}{N}\left( N-2 - 2 i \sqrt{N-1}\right)
\end{equation} then the matrix $P$ assumes the form
\begin{equation}
\left( \begin{array}{ccccccc}
 0 & 0  & 0  & \cdots & 0 & i \sqrt{N-1} & - i \sqrt{N-1} \\
-1 & -1 & -1 & \cdots & -1 & 1 & 1 \\ 0 & 0  & 0  & \cdots  & 1 &
1 & 1 \\ 0 & 0 & 0 & \cdots & 0 & 1 & 1 \\ \vdots & \vdots &
\vdots & \vdots & \vdots & \vdots  & \vdots
\\
0& 0& 1& \cdots & 0&1 &1 \\ 0& 1& 0& \cdots & 0& 1&1 \\ 1& 0 & 0 &
\cdots & 0& 1& 1
\end{array}
 \right).
\end{equation}
Since $|\eta|=1$, we can rewrite the complex number $\eta$ as
$\expon^{-i \theta}$, where, using eq(\ref{rt}),
\begin{equation}
\cos \theta = \frac{N-2}{N} \mbox{\rm ~ ~ and  ~  ~ } \sin \theta
= \frac{2 \sqrt{N-1}}{N}. \label{rot}
\end{equation}

Finally, we note that the probability of finding the search item
can be evaluated to be $\displaystyle \frac{1}{N} (\cos(m \theta)
+ \sqrt{N-1} \sin(m \theta))^2$ while the probability of getting
one of the other entries in the database is $\displaystyle \frac{1}{N}
(\cos(m \theta) - \frac{1}{\sqrt{N-1}} \sin(m \theta))^2$.

\section{Noise induced through decoherence}\label{search}

\subsection{Superoperators} \label{super}

Decoherence can be studied and understood within the context of
superoperators through the evolution of a bipartite quantum
system\cite{preskill2}. A superoperator describes a linear map,
$\$$, from input density matrix, $\rho_{\mbox{\rm in}}$, to output
density matrix, $\rho_{\mbox{\rm out}}$. Although the
superoperator, $\$$, is not a unitary operator, it is a linear map
that preserves hermiticity as well as the trace.

The standard procedure of understanding the behavior of one part
of a bipartite quantum system is to extend the system to a larger
one (in which the environment (E) is incorporated) so that the
evolution of state becomes unitary under the transformation, $\U$.
By assuming complete positivity of the superoperators, it is
possible to study the non-unitarity evolution of the state of a
subsystem using an operator sum representation. In terms of the
operator sum or Kraus representation, we can express this map,
$\$$, more succinctly as \beq \rho_{\mbox{\rm out}} =
\$(\rho_{\mbox{\rm in}}) = \sum_{\mu} M_\mu \rho_{\mbox{\rm in}}
M^\dagger_\mu. \eeq Unitarity of the operator, $\U_{CE}$, also
requires that the Kraus operators satisfy the condition \beq
\sum_\mu M_\mu M_\mu^\dagger = 1 \eeq

\subsection{Decoherence Model}\label{model}

For two-spin-1/2 system, we can consider the
following Kraus  representation \bseq 
M_0 & = & \left( \begin{array}{cc} \sqrt{1 - p} & 0
\\ 0 & \sqrt{1-p} \end{array} \right)\\
M_1 & = & \left( \begin{array}{cc} 0 & \sqrt{p/3}
\\ \sqrt{p/3} & 0 \end{array}\right) =\sqrt{p/3} \sigma^x\\
M_2 & = & \left( \begin{array}{cc}0 & -\sqrt{p/3}
\\ \sqrt{p/3} & 0 \end{array} \right) = \sqrt{p/3} i \sigma^y \\
M_3 & = & \left( \begin{array}{cc} \sqrt{p_3} & 0
\\ 0 & -\sqrt{p_3} \end{array} \right) = \sqrt{p_3} \sigma^z
\eseq 
where $\sigma^i, ~ i = x,y,z$ are the usual Pauli matrices.  We
can check easily that the unitarity condition for the larger space
is satisfied since $\dis \sum_{\mu} M_\mu^\dagger M^{\mu} = 1$.
This decoherence model is sometimes alluded to as the depolarizing
channel.

This decoherence model is an idealization of a transmission and
storage process\cite{adami} with some elegant mathematical
symmetry in which the quantum state in the channel can undergo a
bit-flip and phase errors.  The construction of a depolarizing
channel arises from the interaction of a quantum state,
$|\Psi\rangle$ with an environment in which there is a probability
$p$ that the quantum state will survive and a probability $p/3$
that it would execute a pure bit-flip, a pure phase error or a
combination of both. Thus, schematically, we have \begin{eqnarray}
|\Psi\rangle & \stackrel{1-p}{\longrightarrow} & |\Psi \rangle \\
 |\Psi\rangle &
\stackrel{p/3}{\longrightarrow} & \sigma_1 |\Psi \rangle \\ |
\Psi\rangle & \stackrel{p/3}{\longrightarrow} & \sigma_3 |\Psi
\rangle \\ | \Psi\rangle & \stackrel{p/3}{\longrightarrow} &
\sigma_2 |\Psi \rangle
\end{eqnarray}

Furthermore, it is possible to show that for general spin-1/2
particle in which the density matrix describing the mixed states
is given by $\rho_0 = 1/2 (1 + \vec{n} \cdot \sigma )$, that the
original density matrix $\rho_0$ evolves under the depolarizing
channel as \beq \rho^\prime = p \frac{1}{2} {\bf I_2}  + (1 - p)
\rho_0 \label{evolve1} \eeq where ${\bf I_2}$ is the $2 \times 2$
unit matrix. In order to apply this model to the quantum search
algorithm, it is necessary to consider the extension to $N$
qubits. It is not hard to verify that in this case,
Eq.(\ref{evolve1}) becomes \beq   \rho^\prime = p \frac{1}{N} {\bf
I_N}  + (1 - p) \rho_0 \label{evolve2} \eeq where ${\bf I_N}$ is
the $N \times N$ unit matrix.

\section{Search with decoherence}\label{noisysrch}

\begin{figure}
\epsfig{file=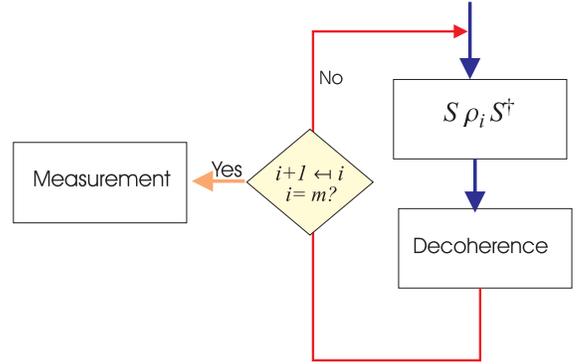, height=.2\textheight} \caption{Schematic
illustration of a decoherence model. }\label{pic}
\end{figure}

Let us now study the effects of the decoherence (depolarizing)
model to the quantum search algorithm. As in any search algorithm,
we first initialize the system into a superposition of states with
equal probabilities. Thus the initial density matrix of the system
is \beq (\rho_0)_{ij} = \frac{1}{N}. \eeq This state is then
subject to the usual inversion-diffusion process so that the
density matrix at the end of the first transformation is \beq
\rho_0^\prime = S \rho_0 S^\dagger. \eeq In the noise-free
Grover's search algorithm, this transformation is repeated a
certain number of times.  In our model, before we perform the
second iteration, we allow the system to evolve under a
depolarizing channel so that using Eq.(\ref{evolve2}), we have
\beq \rho_{i+1} =  p \frac{1}{N} {\bf I_N} + (1 - p) \rho_i.  \eeq
It is easy to show inductively that for $m$ iterations, \bseq
\rho_{m}  & = & p \frac{1}{N} {\bf I_N} \big( 1 + (1-p) + \cdots
\nonumber \\& & + (1-p)^{m-1} \big) + (1 - p)^m S^m \rho_0
S^{\dagger m} \rho_i. \\ & = & \frac{1}{N} \big[ 1 - (1- p)^m
\big] \nonumber \\ & & \mbox{\hspace{2 cm}} + (1-p)^m S^m \rho_0
S^{\dagger m} \label{iterate}\eseq A schematic representation of
the model is shown in Fig. \ref{pic}.

After the $m$ iterations, the state of the system is measured
using the standard Von-Neumann measurement or positive operator
valued measurement through appropriate extension to a higher
dimensional space, so that the state after the measurement reads
\beq \rho_f = \sum_i E_i \rho_m E_i^\dagger \eeq where $E_i$ are
orthogonal projection operators (POVM) $|i\rangle \langle i |$.
However, we note that $\sum_i E_i S^m \rho_0 S^{m \dagger}
E_i^\dagger$ is just the probability of finding the desired state
under the search without decoherence. Moreover, using the results
in section \ref{grover} or  ref. \cite{kwek,pablo}, we have \beq
\sum_i E_i S^m \rho_0 S^{m \dagger} E_i^\dagger = \frac{1}{N}
(\cos m \theta + \sqrt{N-1} \sin m \theta )^2, \eeq where $\dis
\cos \theta = 1 - \frac{2}{N}$ and $\dis \sin \theta = \frac{2
\sqrt{N-1} }{N}$. Finally, using Eq.(\ref{iterate}) and
manipulating using some simple algebra, we find that the
probability of searching successfully for the desired state under
decoherence is \bseq P(m) & = & \frac{1}{N} \big( 1 - (1 - p)^m
\big[ (\cos m \theta + \nonumber
\\ & & \mbox{\hspace{2cm }}\sqrt{N-1} \sin m \theta )^2 + 1
\big]\big) \label{final} \eseq Eq. (\ref{final}) is the main
result in our paper.

For the case of a generalized measurement, we can consider the set
of $r$ POVMs given by $\displaystyle F_i = \sum_{i j} \lambda_{ij}
E_j, ~ i = 1 \cdots r, j = 1 \cdots N $ where the parameters
$\lambda_{ij}$ must satisfy the unitarity condition $\displaystyle
\sum_{i=1}^r \lambda_{ij} = 1$ for each $j$. For
simplicity, we can chose $r=N$ and $\lambda_{ii} = (1 -
\epsilon), ~ \lambda_{ij} = \frac{\epsilon}{r-1}, ~ i\neq
j$.  Indeed, it is not difficult to
imagine an experimental setup for these
POVMs since they can be used to describe possible
errors in the photon detectors.
In this case, we find the probability of successfully
searching for the desired state is \begin{eqnarray} P(m) & = &
(1 - \epsilon) ~ ~ P(m)_{\mbox{\footnotesize ortho}} +
\nonumber
\\ & & \mbox{\hspace{0.2cm }}
\frac{\epsilon}{N} \big( 1 - (1 - p)^m \big[ (\cos m \theta +
\nonumber
\\ & & \mbox{\hspace{2cm }} - \frac{1}{\sqrt{N-1}} \sin m \theta )^2 + 1
\big]\big)\label{povm}
\end{eqnarray}
where $P(m)_{\mbox{\footnotesize ortho}}$ is the probability of
the search item under Von Neumann orthogonal measurements and
given in Eq.(\ref{final}).

\section{Analysis}\label{discuss}

Within the search algorithm, the number of iterations needed for
the search is usually fixed by maximizing the probability of
getting the desired state. Without any decoherence, the number of
iterations $m_{\mbox{\rm max}}$ of the first maximum for large $N$
is found to be \beq m_{\mbox{\rm max}} = \frac{\pi \sqrt{N}}{4}.
\eeq With decoherence, from Eq.(\ref{final}), the condition is
imposed by the following transcendental equation, \beq \frac{m}{1
- p} \big( 1 - \frac{N}{2} (1 - \cos \zeta)\big) = \frac{N
\zeta}{2 m + 1} \sin \zeta \eeq where $\zeta = 2 (2 m+1) \theta$.

\begin{figure}
\epsfig{file=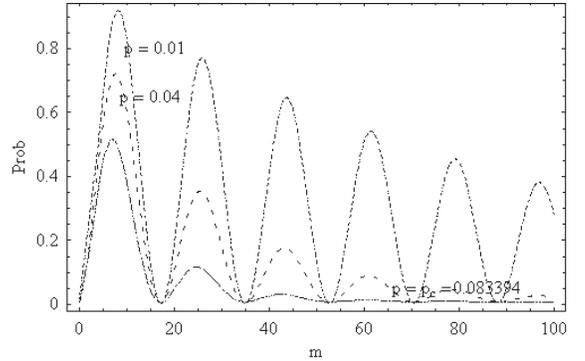, height=.2\textheight} \caption{Probability
of search item as a function of the number of iterations for
$p=0.1, 0.4$ and $0.83394$ and $N=128$. }\label{fig2}
\end{figure}

It is interesting to compare the probability of the search item as
a function of the number of iterations, $m$, for the specific case
for various values of $p$. Such a graph for $N=128$ and $p=0.01,
0.04$ and $0.083394$ is shown in Fig. \ref{fig2}.  The value of
$p=p_c= 0.083394$ refers to the maximum allowed value of the
decoherence parameter $p$ subject to $m_{\mbox{\rm max}} = \pi
\sqrt{N}/4$ such that the probability of the search item do not
fall below 0.5. Fig. \ref{fig2} clearly shows that the presence of
decoherence tends to provide a decaying effect on the periodic
probability distribution of the search item. In the noise-free
situation, we have iterations in which the probability of search
items reaches a periodic maximum without significant decay in the
maximum probability. In the case of a noisy search, we see that
there is gradual reduction in the maxima so that subsequent maxima
are usually rendered useless in the search as the magnitude of
decoherence gets larger.

\begin{figure}
\epsfig{file=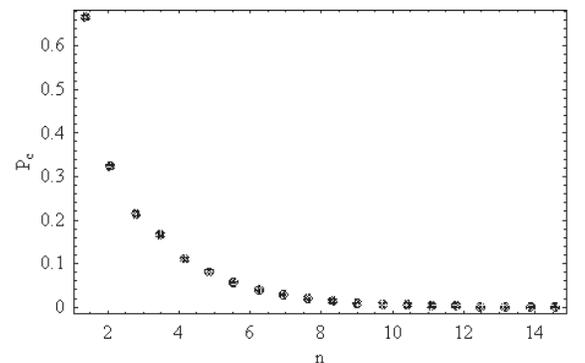, height=.2\textheight} \caption{Critical
decoherence parameter $p_c$ as a function of $n =
log_2(N)$}\label{fig1}
\end{figure}

As we have noted earlier, if we maintain the number of iterations
to the value required under the noise-free situation, namely
$m_{\mbox{\rm max}}$, we can solve Eq.(\ref{final}) for the value
of $p_c$ such that the probability of the search item is above
chance level (i.e. $P(m) > 0.5$) for $p < p_c$. In Fig \ref{fig1},
we plot the critical decoherence parameter $p$ as a function of
the database size $n= log_2(N)$ up to a database size of about $2
\times 10^6$. This graph shows that as the database gets larger,
it is imperative and necessary to overcome the problem of
decoherence. At a database size of $2^{21} \approx 2 \times 10^6$,
the amount of decoherence allowed is only about $0.000609$, which
is a very small value indeed.

\begin{figure}
\epsfig{file=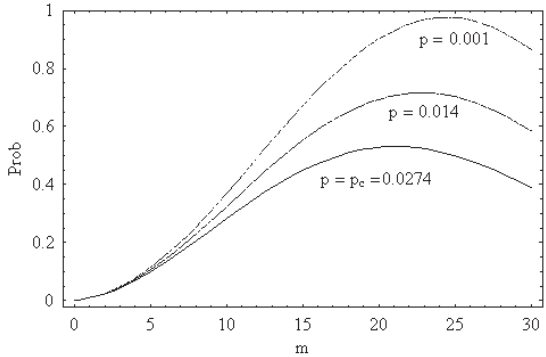, height=.2\textheight} \caption{Probability
of the search item as a function of the number of iterations $m$
for  $p= 0.001, 0.014$ and $p=p_c= 0.0274$ }\label{fig3}
\end{figure}

\begin{figure}
\epsfig{file=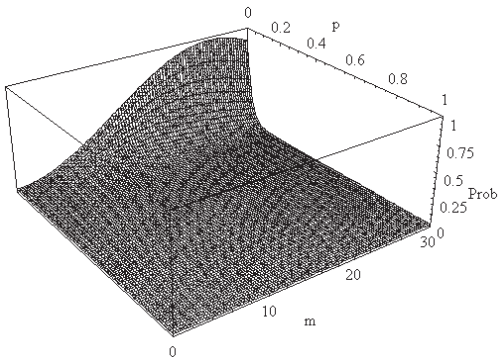, height=.2\textheight}
\caption{Three-dimensional plot of the probability of the search
item as a function of the number of iterations, $m$ and
decoherence parameter, $p$ ($N=1024$) }\label{fig4}
\end{figure}

\begin{figure}
\epsfig{file=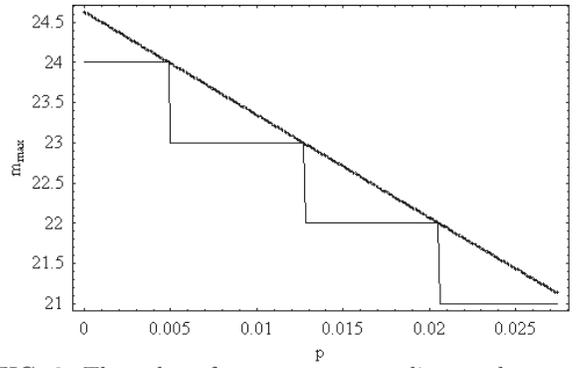, height=.2\textheight} \caption{The value of
$m_{\mbox{\rm max}}$ corresponding to the maximum probability of
the search item as a function of $p$ }\label{fig5}
\end{figure}

In a general search problem, the number of iterations can always
be controlled. For a noisy search algorithm, it seems from our
analysis that it may be better to reduce the number of iterations
so that the probability of the search items becomes a little
larger. Using a database of $N=1024$, we plot the variation of the
probability of the search items as a function of iterations $m$
for $p= 0.001, 0.014$ and $p=p_c= 0.0274$ in Fig. \ref{fig3}. The
graph clearly shows a shift in the maxima for the curves. Indeed
for the near critical value of $p= p_c$, we see that the maxima
hovers around the value of 0.5 at $m=21$. A three-dimensional plot
of the probability as a function of $m$ and $p$ is shown in Fig
\ref{fig4}. Indeed, by varying the decoherence parameter $p$, it
is possible to solve for the value of $m_{\mbox{\rm max}}$
corresponding to the probability of the search item. Fig.
\ref{fig5} shows the value of $m_{\mbox{\rm max}}$ as a function
of $p$. The step function refers to the integer part of
$m_{\mbox{\rm max}}$.

\begin{figure}
\epsfig{file=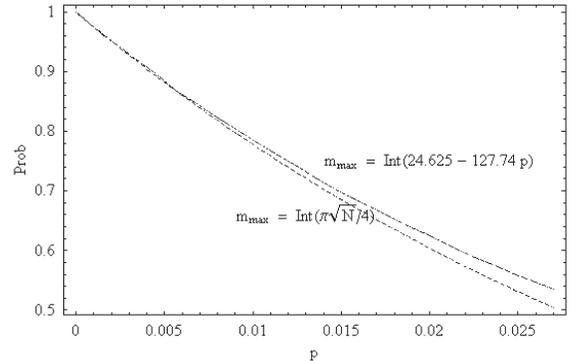, height=.2\textheight} \caption{Probability
of the search item as a function of $p$ for different iteration
formulae. }\label{fig6}
\end{figure}

It is interesting to note that the graph of $m_{\mbox{\rm max}}$
corresponding to the maximum probability of the search item is a
linear function of $p$. A interpolation fit gives $m_{\mbox{\rm
max}} = 24.6254- 127.7426 p$. If we then plot the probability of
the search item as a function of $p$ using the integer part of
this linear function, $i.e.{\rm  Int}(m_{\mbox{\rm max}})$ and
compare it with the corresponding probability for $m_{\mbox{\rm
max}}={\rm  Int}(\pi \sqrt{N}/4)$, we see that the difference is
reasonably negligible, especially for small $p$. This plot is
illustrated in Fig. \ref{fig6}.

\begin{figure}
\epsfig{file=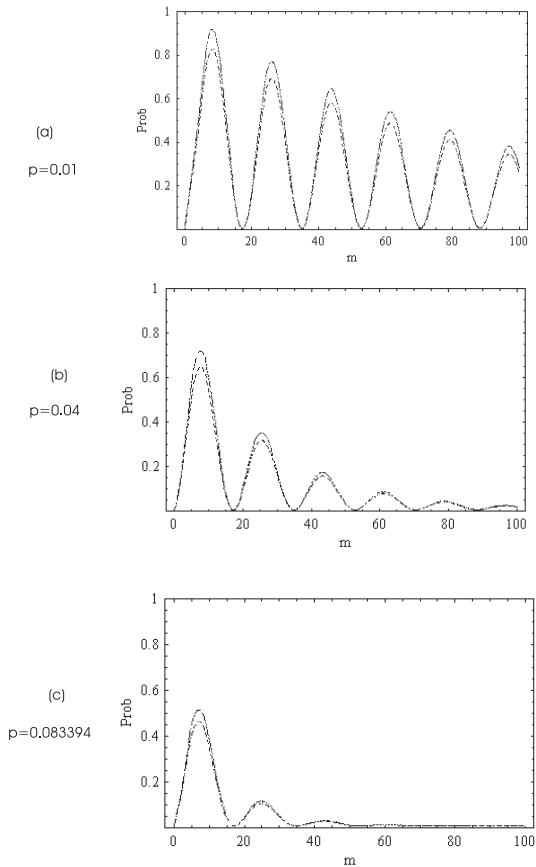, height=.48\textheight}
\caption{Probability of the search item as a function of $m$ for
$N=128$, $\epsilon=0.1$ and (a) $p=0.01$, (b) $p= 0.04$ and (c) $p = 0.083394$ for
orthogonal (bold) and non-orthogonal (dotted)
measurements. }\label{povm1}
\end{figure}

In the context of a generalized measurement, we find that
probability amplitude of the search item is given by the
expression in Eq.(\ref{povm}) and this probability generally
reduces in the first maximal iteration. The plots of the
probability of the search item as a function of the number of
iterations $m$ for various values of $p$ at $N=128$ and $r=131$ is
plotted in Fig. \ref{povm1}. In the plots, we have chosen a reasonably
small value for the parameter, $\epsilon =0.1$, since under
experimental conditions, one do not expect a large value for this
parameter. In fact, the results show that
for sufficiently small $\epsilon$, it is possible to
confine our study to orthogonal Von Neumann measurements.

\section{Discussion and Concluding Remarks}\label{conclude}

In this paper, we study the effects of the search algorithm
analytically and numerically using a simple decoherence model.
Several observations are in order.  Firstly, we see that the
search algorithm can be rendered useless for small degree of
decoherence under the depolarizing channel as the database gets
larger. Indeed for a database size of about 2 million entries, the
degree of decoherence, $p$, is a factor of 1000 smaller than the
corresponding value for a database size of 4. It appears that for
large $N$, the factor varies approximately as $\sqrt{N}$.

Secondly, as we would expected, decoherence generally diminishes
the probability of the search items.  With high decoherence, it
may not be useful to iterate beyond the first maxima in the
probability distribution of the search item.

Thirdly, the number of iterations needed to maximize a search
probability decreases with increasing decoherence. This is a
somewhat unexpected result since decoherence ostensibly tends to
reduces the probability distribution. Nevertheless, although the
associated maximum probability has decreased, it is interesting to
note that under a noisy channel, one could in general improve the
efficiency by reducing the iterations prior to measurement. It is
interesting to note here that the number of iterations (considered
as a continuous variable, i.e. prior to taking the integer part)
falls off linearly with the degree of decoherence, $p$. However,
if we continue to use the maximum number of iterations in
noise-free situation, the difference in probability appears to be
less than 10\%.

Finally, we note that although the decoherence model that we have
employed here is very simple, it certainly provides a wealth of
information regarding the behavior of the search algorithm under a
noisy environment.

\end{multicols}
\end{document}